\documentclass[proof]{pasj01}

\begin{document}
\SetRunningHead{S. Nagatomo et al.}{Extinction Law toward Galactic Center}

\title{Interstellar Extinction Law toward the Galactic Center IV: 
J, H, and Ks Bands from VVV Red Clump Stars}


%
\author{%
   Schun \textsc{Nagatomo}\altaffilmark{1},
   Tetsuya \textsc{Nagata}\altaffilmark{1},
   and
   Shogo \textsc{Nishiyama}\altaffilmark{2}} %
 \altaffiltext{1}{Department of Astronomy, 
Graduate School of Science, 
Kyoto University, Kyoto 606-8502}
 \altaffiltext{2}{Miyagi University of Education, Aoba-ku, Sendai 980-0845, Japan}
 \email{nagatomo@kusastro.kyoto-u.ac.jp}


\maketitle

\begin{abstract}
We have determined the wavelength dependence of the extinction in the near-infrared bands 
($J$, $H$, $K_{\mathrm S}$) toward the Galactic center from 
the VVV 
(VISTA Variables in the V\'ia L\'actea) 
aperture photometry of the stars in the region 
$|l| \lesssim 2\arcdeg.0$ and $0\arcdeg.5 \lesssim |b| \lesssim 1\arcdeg.0$; 
this region consists of 12 VVV tiles.  
We have found significant systematic discrepancy 
up to $\sim 0.1$ mag 
between the stellar magnitudes 
of the same stars in overlapping VVV tiles.  
However, 
by carefully using the positions of red clump stars in color-magnitude diagrams 
as a tracer of the extinction and reddening, 
we are able to determine the average of the ratios of total to selective extinction to be 
$A(K_{\mathrm S}) / E(H-K_{\mathrm S}) = 1.44 \pm 0.04 $,
$A(K_{\mathrm S}) / E(J-K_{\mathrm S}) = 0.423 \pm 0.024$,
$A(H)             / E(J-H)             = 1.25 \pm 0.04$; 
%
%
from these ratios, a steep power law 
$A(\lambda) \propto \lambda^{-\alpha}$ 
whose index $\alpha$ is $\sim 2.0-2.3$ in the 
$J, H, K_{\mathrm S}$ wavelength range is estimated.  
The obtained wavelength dependence is consistent with those obtained 
with the Mauna Kea Observatory (MKO) photometric system employed in 
the Simultaneous 3-color InfraRed Imager for Unbiased Survey 
(SIRIUS) camera attached to the Infrared Survey Facility (IRSF) telescope 
in previous studies.   
Such a steep decline of extinction toward the longer wavelengths 
is also in line with recent results based on deep imaging surveys.  
The determined extinction law seems to be variable in the VVV tile to tile,  
and it is not clear how much of this is due to real sight line variations and 
due to observational systematic effects.  
Thus, there might be room for improvement of the extinction law determination 
from the existing VVV data, but 
this steep extinction law tends to locate heavily reddened objects in the Galactic plane 
more distant from us when their distance moduli are calculated from the observed reddening values.  
\end{abstract}
\section{Introduction}
The clump of red giants (hereafter red clump, RC) is a striking feature 
in the color-magnitude diagrams (CMDs) of the 
bulge
of the Milky Way Galaxy.   
The near constancy of intrinsic properties of the RC stars 
are now widely recognized, and 
they are employed in the ``RC method'', in which 
candidate RC stars are used as probes of interstellar extinctions 
and other quantities \citep{gir16}.  
Following the determination of 
the total to selective extinction 
$A(V) / E(V-I)$ 
by \citet{woz96}
and 
$A(I) / E(V-I)$ 
by \citet{uda03}, 
Nishiyama et al. (2006; hereafter N06) determined the 
wavelength dependence of the extinction in the near-infrared bands 
($J$, $H$, $K_{\mathrm S}$) toward the Galactic center 
from the observation of RC stars with 
the 
Mauna Kea Observatory
(MKO) filters \citep{tok02} employed in 
the Simultaneous 3-color InfraRed Imager for Unbiased Survey
(SIRIUS) camera 
attached to the 
Infrared Survey Facility
(IRSF) telescope \citep{nag03}.  

%
Nishiyama et al. (2009; hereafter N09) examined the 
wavelength dependence of the extinction from 1.2 to 8.0$\mu {\mathrm m}$ using 
the Two Micron All Sky Survey (2MASS) and Spitzer/IRAC/GLIMPSE II catalogs.  
However, 
these two catalogs are not deep enough to contain all the RC stars.  
To overcome the limitation, N09 assumed that the center of distribution in the lines of sight is at the same distance from us both for
the RC giants and the giants in the upper red giant branch (RGB).  
They then determined the reddening 
$E(K_{\mathrm S}-\lambda)$ 
of the upper RGB for the 2MASS and IRAC wavebands 
while using the IRSF/SIRIUS determination of the extinction 
$A(K_{\mathrm S})$ 
of the RC stars.
The derived wavelength dependence of the extinction in the  
2MASS $J$, $H$, and $K_{\mathrm S}$ bands showed good agreement with the MKO system.  
N09 have come to the conclusion that the extinction  
is well fitted by a power law of steep decrease
$A(\lambda) \propto \lambda^{-2.0}$
toward the GC, 
in contrast to the 
\citet{car89} fit to the \citet{rie85} result of gentle decrease 
$A(\lambda) \propto \lambda^{-1.6}$ in the near infrared.  

%
However, there seems to be a slight difference between the derived wavelength dependences 
from the IRSF/SIRIUS MKO photometric system (with the ``pure'' RC method) and 
the 2MASS system  (with the RC method variant mentioned above).  
In particular, the total to selective extinction ratio 
$A(K_{\mathrm S}) / E(H-K_{\mathrm S})$,  
probably most useful in determining the extinction from the observed reddening 
of highly reddened stars, 
is $1.44 \pm 0.01$ 
in the IRSF/SIRIUS MKO system, but 
it 
is calculated 
to be $1.66 \pm 0.06$ 
if the 
$A(H)$ and  
$A(K_{\mathrm S})$ 
values for 
the 2MASS system 
in Table 1 of 
N09 are used\footnote{
\citet{sch14} used a ratio 
$A(K_{\mathrm S}) / E(H-K_{\mathrm S}) = 1.61$ 
from the same table by 
N09, 
but this is based on the observation of 2MASS $H$ and IRSF/SIRIUS $K_{\mathrm S}$.}.  
\citet{dek15a} transformed this larger ratio in the 2MASS system 
to the VISTA system \citep{sai12} 
and 
obtained
$A(K_{\mathrm S}) / E(H-K_{\mathrm S}) = 1.63$, 
which is then used to determine
extinctions and distances of classical Cephids they discovered
in the 
VISTA Variables in the V\'ia L\'actea 
(VVV) 
survey \citep{min10}.  
This large ratio implies rather gentle decrease of extinction toward the longer wavelength, 
although not as gentle as the 
\citet{car89} curve of 
$A(K) / E(H-K) = 1.83$.  

%
Such a difference can cause a serious discrepancy in determining the distance of objects 
because 
the observed color of a reddened object leads to 
a larger extinction and a smaller distance 
when a larger ratio (gentler extinction decrease) is adopted.  
As long as the target objects are not so much reddened by significant dust, 
the choice of extinction curves does not matter as in the case of type II 
Cepheids in the VVV survey \citep{bha17}, 
where only a small fraction of them suffers 
reddening of $E(J-K_{\mathrm S}) > 1$.  
However, in the case of classical Cepheids, 
most of them are concentrated to the Galactic plane and suffer heavy  extinction;  
\citet{mat16} questioned the distances of the classical Cepheids determined by 
D\'ek\'any et al.(2016a, 2015b), suggested that all their Cepheids are in fact more distant, 
and disapproved the presence of a putative inner thin disk of young stars represented by Cepheids.  

%
%

%
Since the fourth data release (DR4) of the VVV survey is already made public and 
its photometric catalogue based on aperture photometry (see \cite{sai12}) is widely used, 
straightforward determination of the extinction law using these data seems highly desirable.
In this paper, we have determined the ratios of total to selective extinction 
$A(K_{\mathrm S}) / E(H-K_{\mathrm S})$,
$A(K_{\mathrm S}) / E(J-K_{\mathrm S})$, 
and 
$A(H)             / E(J-H)$ 
by simply adopting the RC method on the VVV data 
toward the field observed and analyzed with the RC method by 
N06, 
without the assumption of the coexistence of upper RGB stars and RC stars 
or any transformation between photometric systems.  
In the process of selecting the RC stars, 
we have found that there is significant systematic discrepancy 
up to $\sim 0.1$ mag 
in the stellar magnitudes 
between the aperture photometries of a star in the overlapping VVV tiles.  
However, averaging the ratio in each tile, we have derived the overall extinction tendency 
toward the GC.  
%
%
\section{VVV Data and CMD Analysis}
\begin{figure}
  \begin{center}
    \FigureFile(150mm,150mm){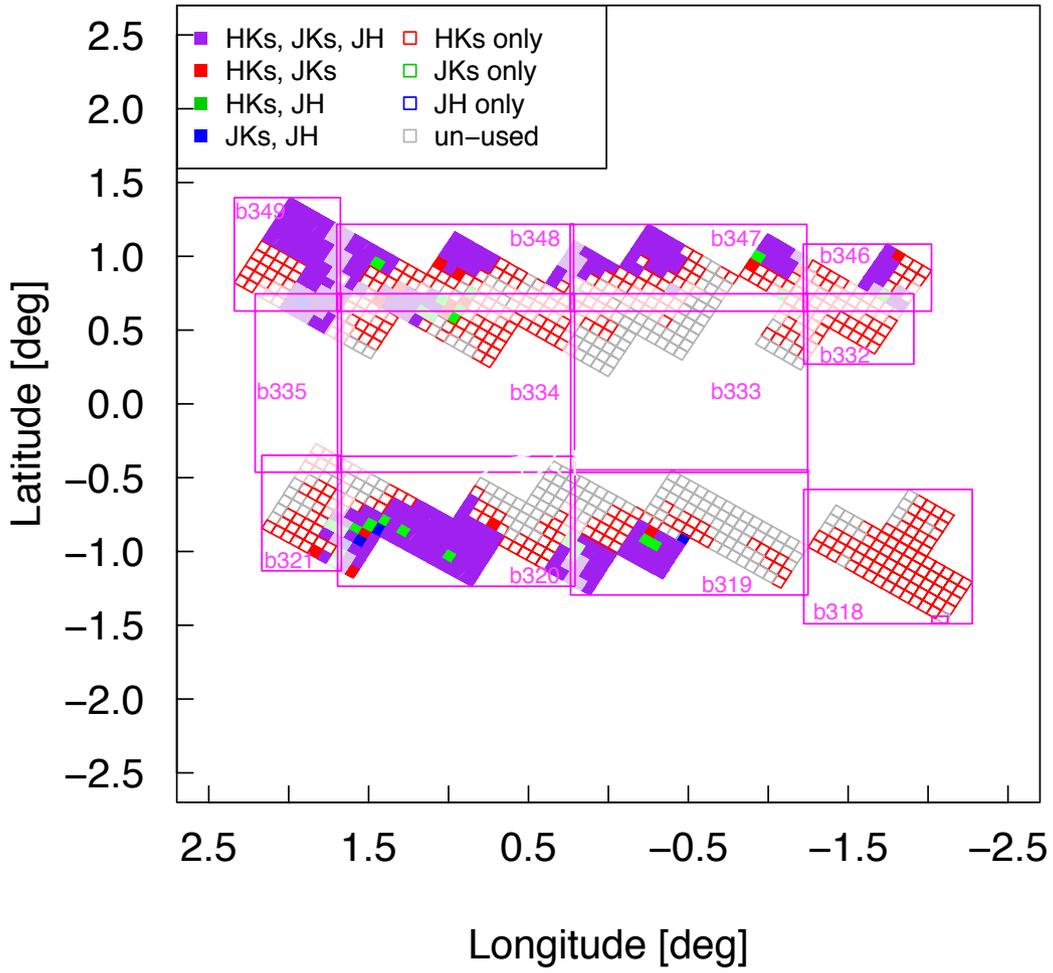}
  \end{center}
\caption{VVV observation area 
and fields used for data analysis. 
The pink rectangle lines are the edges of the 12 VVV tiles in our field analyzed. 
Following N06, each small square is a subfield of $4' \times 4'$.  
Only the colored subfields were used for the data analysis in this paper; 
the gray line squares are the regions where the magnitude and color of RC stars were not reliably determined due to large extinction.  
  }
\label{fig:ObsRegionTile}
\end{figure}

\begin{figure}
  \begin{center}
    \FigureFile(150mm,150mm){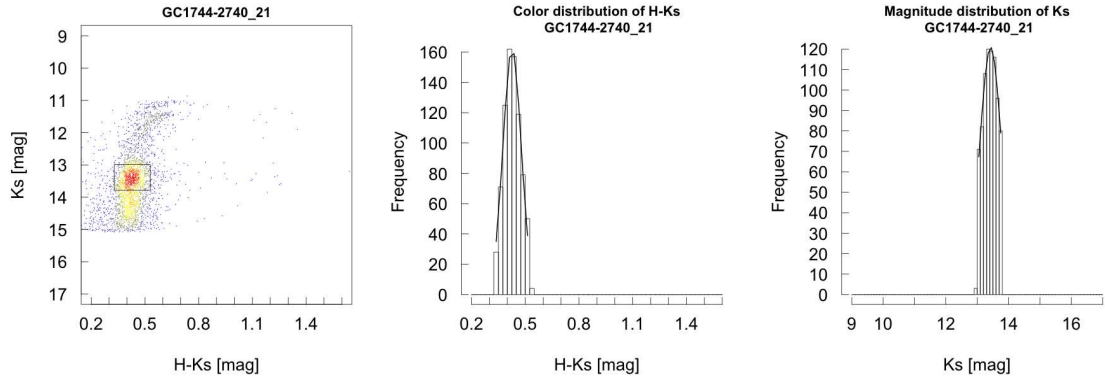}
  \end{center}
\caption{An example of  $K_{\mathrm S}$ vs. $H-K_{\mathrm S}$ CMD of a subfield (left).  
 In the rectangular region of the left panel, a color peak (middle) and a magnitude peak (right) are determined.}
\label{fig:detPeak}
\end{figure}

%
We use the VVV DR4 photometric catalogues\footnote{http://www.eso.org/rm/api/v1/public/releaseDescriptions/80} 
which contain calibrated aperture photometry,  
adopting the ``aperture 1'' magnitudes 
with the aperture diameter of 1.0 arcsec (smallest).  
We determine the ratios of total to selective extinction in the same field as 
N06.  
We might be able to exploit the merits of the VVV data and examine a larger field, 
but we would like to examine 
exactly the same field as 
N06 
to check the possible differences 
(due to sites, telescopes, IR imagers, detectors, and filters; see also \citet{sot13}) 
between the IRSF and VVV surveys.  
First, we divide each square field of $20' \times 20'$ 
in Figure 1 of 
N06 
 (41 in total) 
of $|l| \lesssim 2\arcdeg.0$ and $0\arcdeg.5 \lesssim |b| \lesssim 1\arcdeg.0$, 
into 25 subfields of $4' \times 4'$.  
The VVV data for these fields are from the 12 ``tiles'' b318-321, b332-335, and b346-349 
(Figure \ref{fig:ObsRegionTile}; see also Fig.2 of \citet{sai12} for the designation).  
Then we construct 
$K_{\mathrm S}$ versus $H - K_{\mathrm S}$ CMDs.  
For selective extinction related with the $J$ band,  
we use 18 square fields in total and construct 
$K_{\mathrm S}$ versus $J - K_{\mathrm S}$, and 
$H$             versus $J - H$
CMDs.  
Second, we extract
stars in the region of CMDs dominated by RC stars
(the rectangular region of the CMD in Figure \ref{fig:detPeak}), 
and the extracted stars are used to make magnitude 
and color 
histograms.  
These histograms have 
clear peaks, which are then fitted with a Gaussian function.

%
Due to highly nonuniform interstellar extinction over the whole area 
of $|l| \lesssim 2\arcdeg.0$ and $0\arcdeg.5 \lesssim |b| \lesssim 1\arcdeg.0$, 
the RC peaks in CMDs shift from one sight line to another.
The peaks shift in the range $13.4 \lesssim K_{\mathrm S} \lesssim 14.6$ 
and $0.4 \lesssim H-K_{\mathrm S} \lesssim 1.2$, 
and therefore we have to shift the CMD region to extract
RC stars from subfield to subfield.  
Following 
N06, 
we use only 
the subfields in which the peak magnitude of RC stars is about more
than 0.5 mag brighter than the limiting magnitudes, to 
avoid problems 
of unreliable estimates in too reddened fields.

\section{Results}

%
\begin{figure}
  \begin{center}
    \FigureFile(50mm,100mm){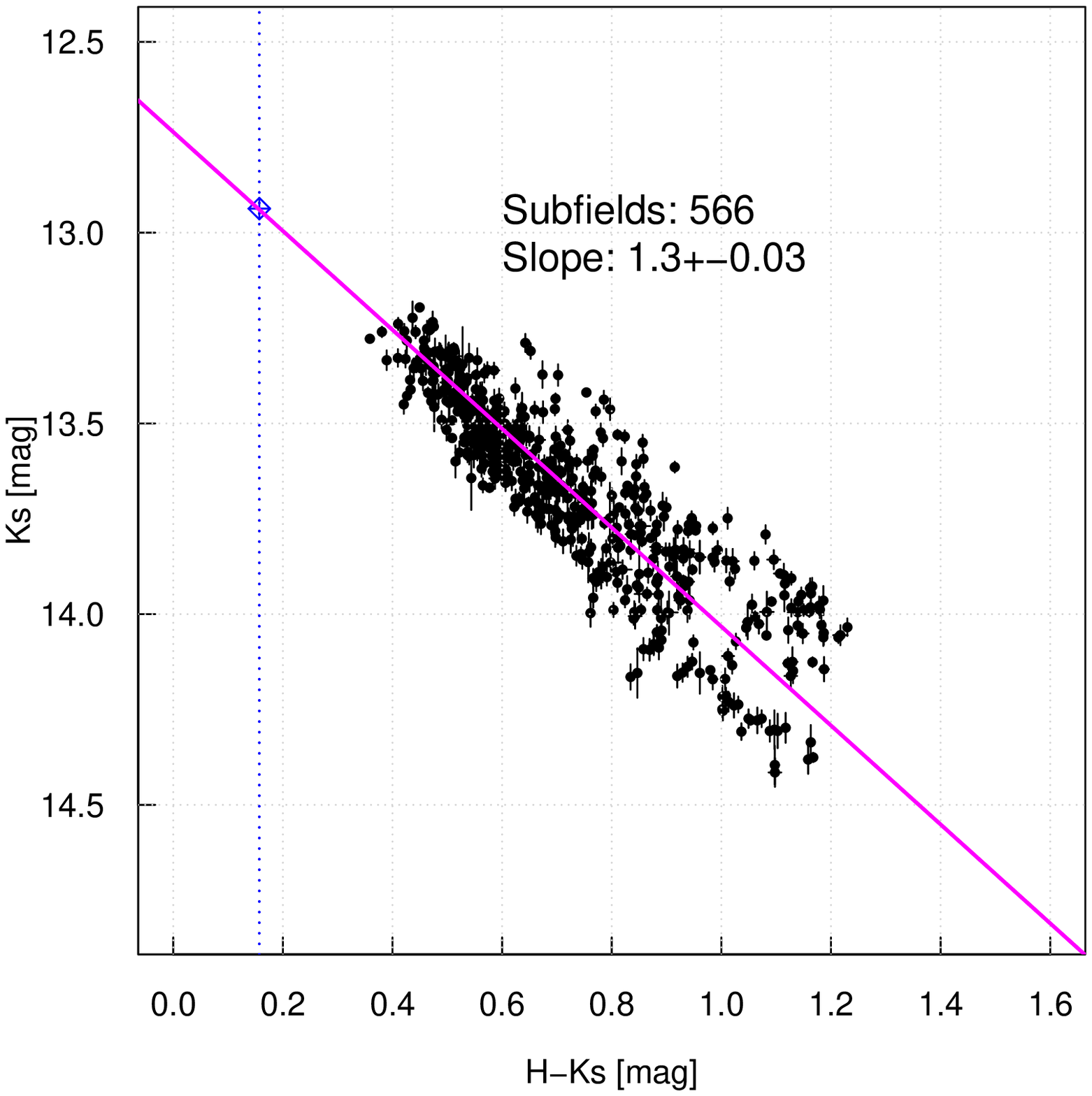}
    \FigureFile(50mm,100mm){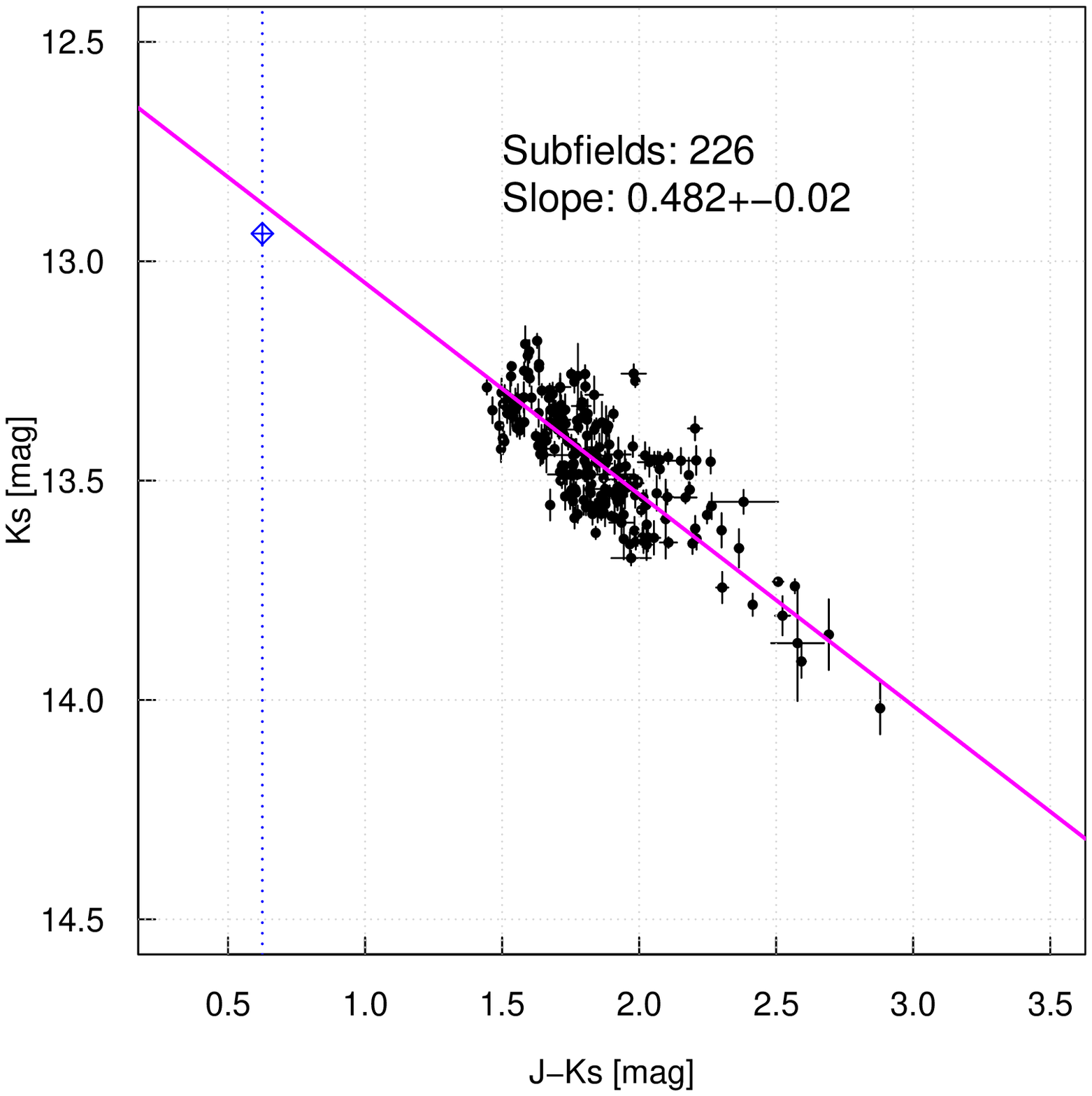}
    \FigureFile(50mm,100mm){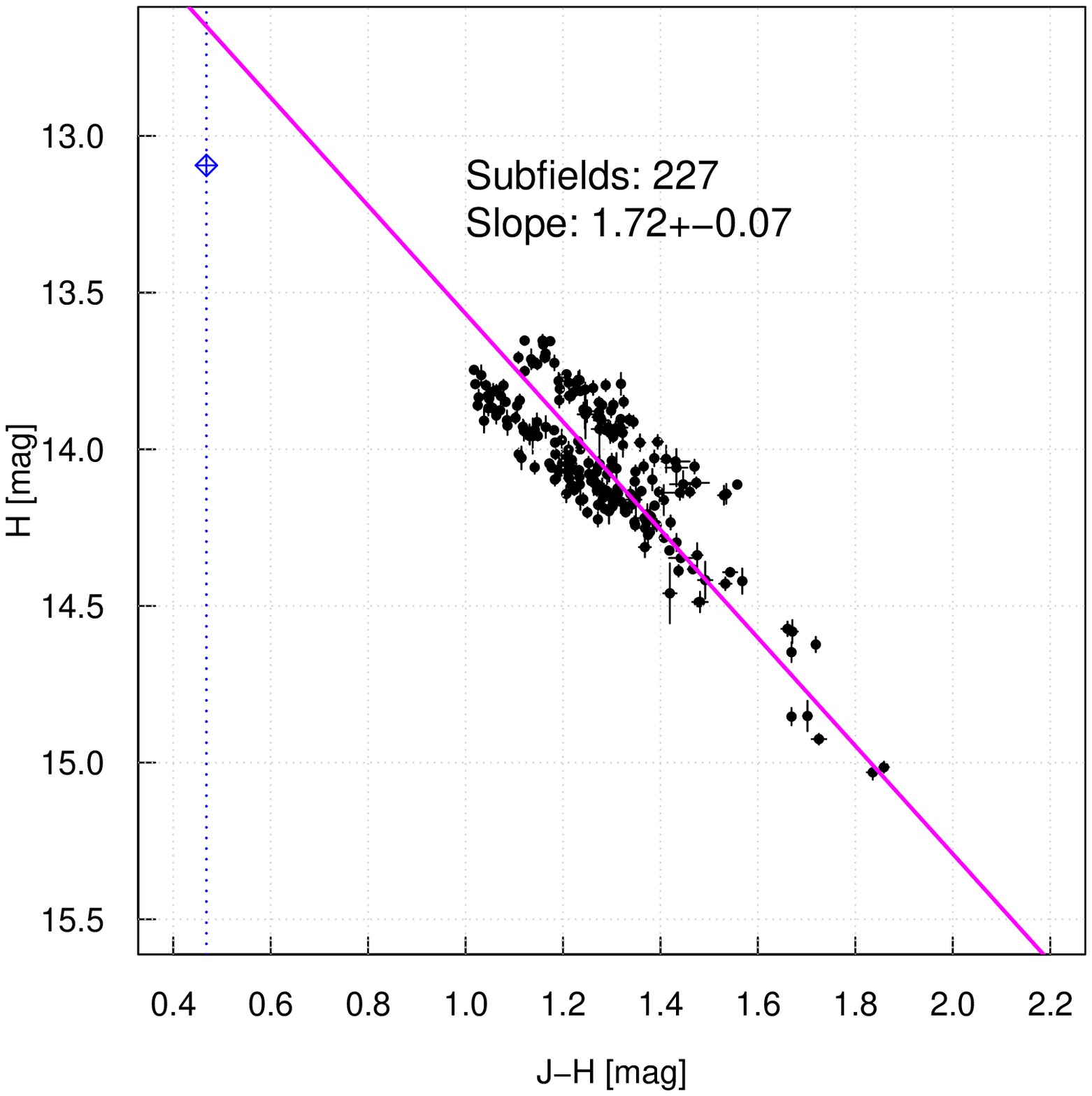}
  \end{center}
\caption{ For the whole data, each magnitude and color Gaussian-fit peaks of RC stars 
in each subfield
 $K_{\mathrm S}$ vs. $H-K_{\mathrm S}$ (left), 
 $K_{\mathrm S}$ vs. $J-K_{\mathrm S}$ (middle), and 
 $H$ vs. $J-H$ (right)
CMDs are shown with the black dots, 
and the least-squares fits to them (pink lines).  
The blue squares at the upper left side represent the predicted apparent magnitudes and intrinsic colors of RC stars free from extinction, following \citet{nat16}.}
\label{fig:slope}
\end{figure}

%


%
Figures \ref{fig:slope}(left), (middle), and (right) present the 
$K_{\mathrm S}$ versus $H - K_{\mathrm S}$,  
$K_{\mathrm S}$ versus $J - K_{\mathrm S}$, and 
$H$             versus $J - H$ 
CMDs 
for the whole data tiles.  
In the figure, each pair of magnitude and color was obtained from each subfield.
The solid lines are 
least-squares fits to the data. 
Slopes of the fitted lines are listed in the second column of Table \ref{tab:slope2}. 
The $A(K_{\mathrm S}) / E(J-K_{\mathrm S})$ slope 
 (middle; 0.482 $\pm$ 0.02) is very similar to 
that obtained by 
N06 (see Table1 bottom), 
but
the $A(H)             / E(J-H) $ slope
 (right; 1.72  $\pm$ 0.07) is rather large, and   
the $A(K_{\mathrm S}) / E(H-K_{\mathrm S}) $ slope
 (left; 1.30  $\pm$ 0.03) is much smaller. 
Here, the slopes are derived from the principal component analysis.  
The original CMD is transformed into the coordinates 
of the first component and the second component, and then 
the possible error from the horizontal axis 
is calculated with the least-squares method.  
Finally, the transformation of the error back to the original color-magnitude coordinates 
produces the error in the slope.  

%

\begin{figure}
  \begin{center}
    \FigureFile(160mm,200mm){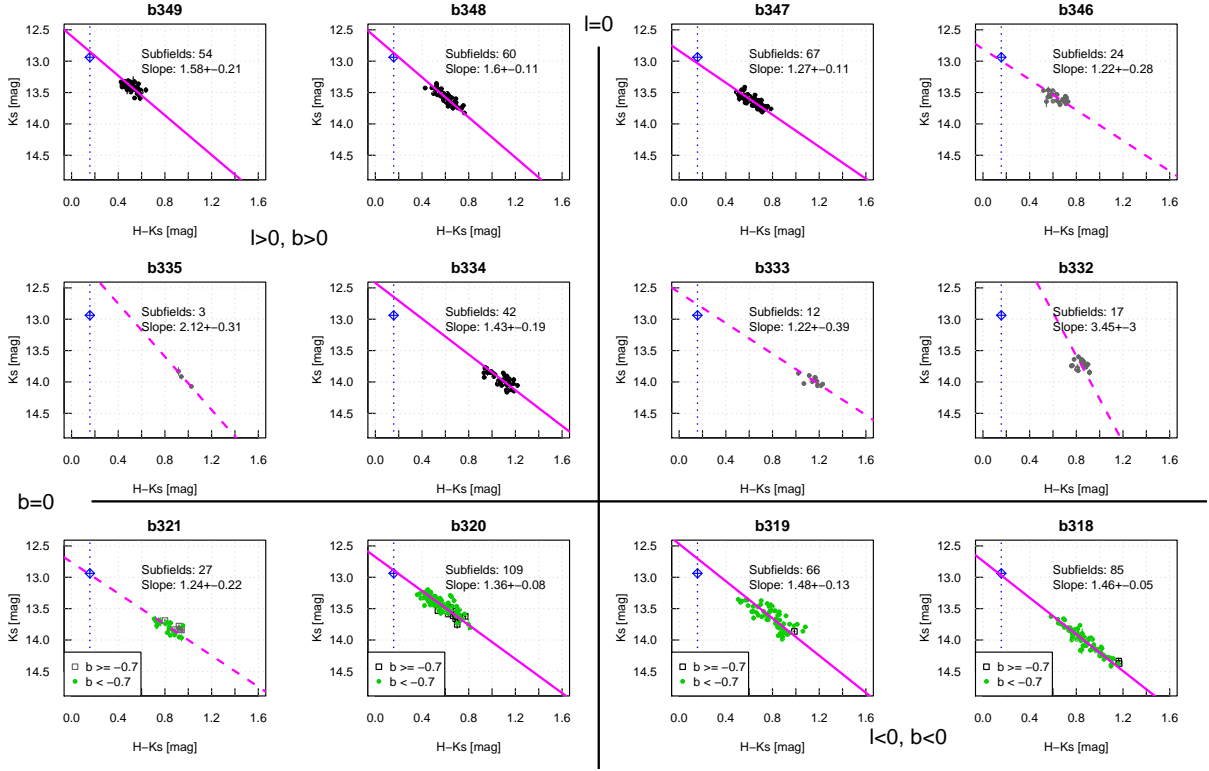}
  \end{center}
\caption{Each magnitude and color Gaussian-fit peaks of RC stars 
in the
$K_\mathrm{S}$ vs. $H-K_\mathrm{S}$ CMDs for the VVV tiles 
$b > 0.7\arcdeg$(top), $ 0.7\arcdeg > b > 0\arcdeg$(middle), and  $ 0\arcdeg > b$ (bottom).  
In the bottom diagrams, 
black squares are  $0\arcdeg > b >-0.7\arcdeg$, and 
green dots are $-0.7\arcdeg > b $ data points.  
The pink lines are the least-squares fits to the data
(Broken lines are poor fits, 
where less than 5\% of total subfields are included in the tile.  
Their slope errors are significant in general).  
The upper left blue squares are 
the predicted apparent $K_\mathrm{S}$ magnitude and intrinsic $H-K_\mathrm{S}$ color 
of RC stars free from extinction.
}
\label{fig:slope2}
\end{figure}

\begin{figure}
  \begin{center}
    \FigureFile(160mm,200mm){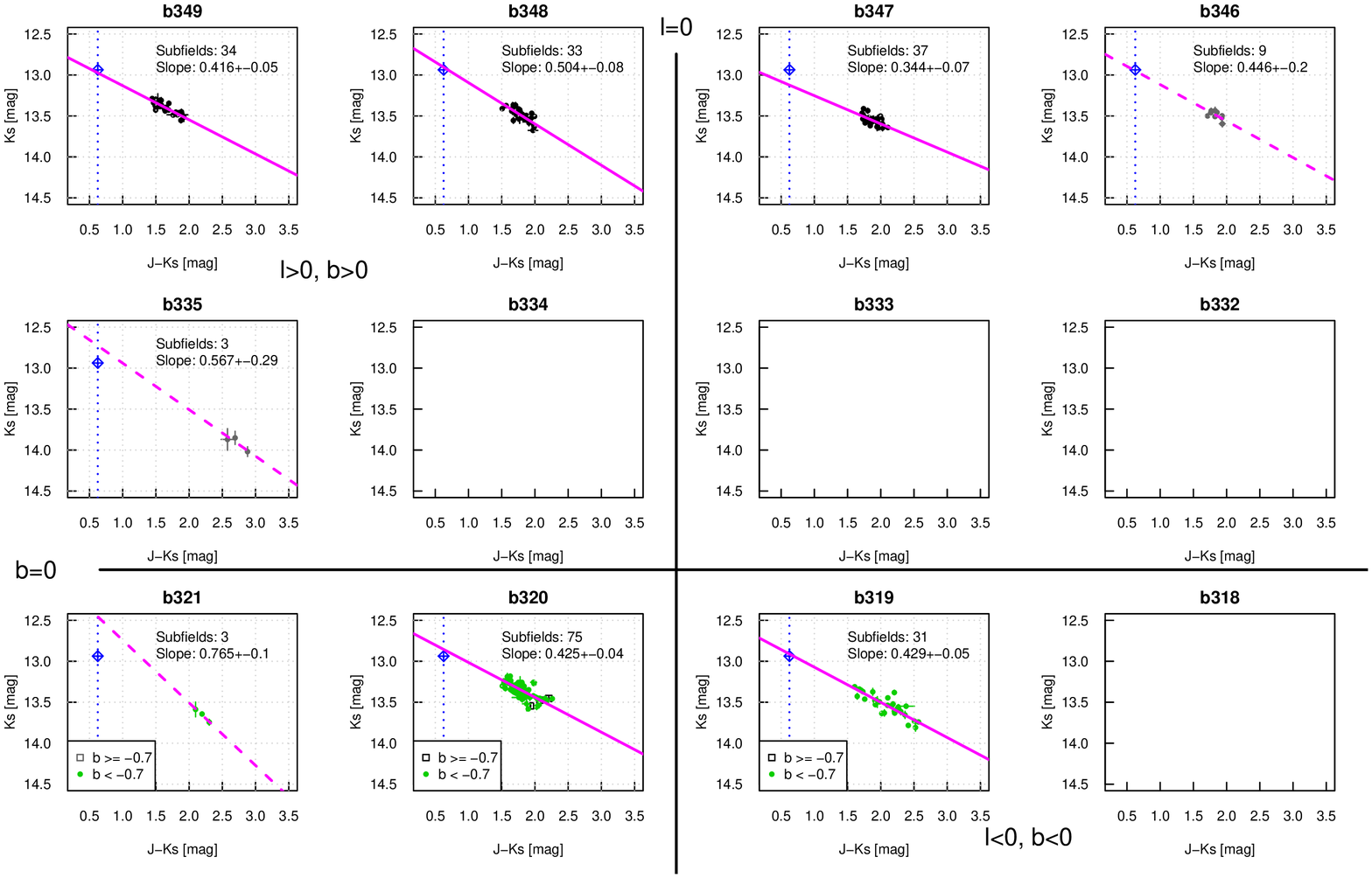}
  \end{center}
\caption{Same as Figure 4, but for $K_\mathrm{S}$ and $J-K_\mathrm{S}$.  
The tiles b332-334 and b318 have no subfields 
where reliable RC peaks can be determined.  

}
\label{fig:slopes2JK}
\end{figure}

\begin{figure}
  \begin{center}
    \FigureFile(160mm,200mm){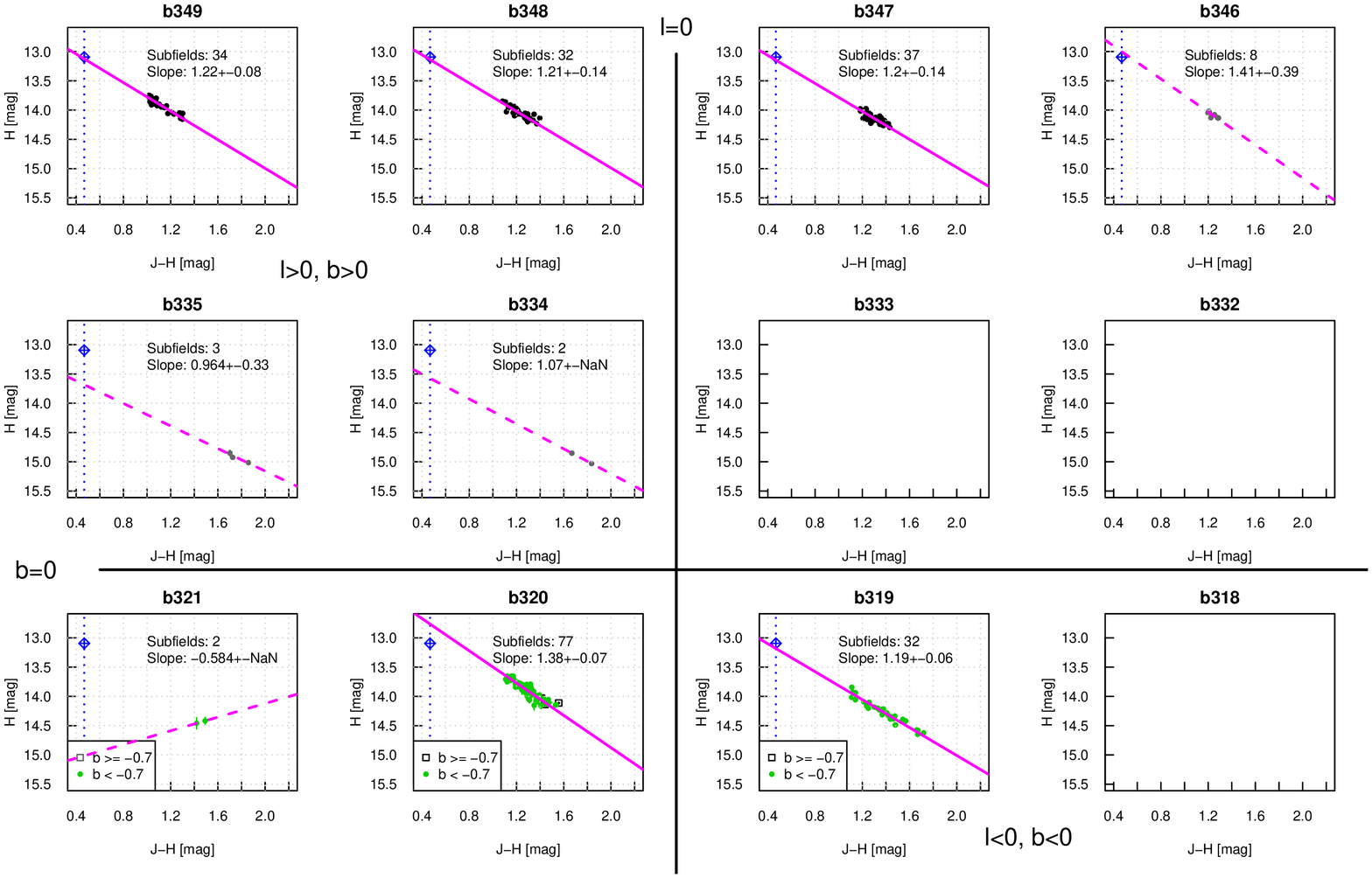}
  \end{center}
\caption{Same as Figure 4, but for $H$ and $J-H$.
The tiles b332, 333, and b318 have no subfields 
where reliable RC peaks can be determined.  
}
\label{fig:slopes2JH}
\end{figure}

We notice that the data in these CMDs are significantly more scattered 
in comparison with those in 
N06.  
Also, the data points seem to be in a few streaks; 
in particular, the $H$ vs. $J-H$ CMD seems to be composed of two or three streaks 
nearly parallel to, but rather gentler than the fitted line, and 
the least-squares fit to the data (pink line) passes far above the upper left blue square 
(the predicted apparent $H$ magnitude and intrinsic $J-H$ color 
of RC stars free from extinction; see below)
because of the several data points in the lower right side.  
The lower part of the $K_{\mathrm S}$ vs. $H-K_{\mathrm S}$ CMD also seems to be split.  
Since the data are taken from the twelve VVV tiles, 
we plot, 
for each tile, 
the $K_{\mathrm S}$ vs. $H-K_{\mathrm S}$
CMD in Figure \ref{fig:slope2}, 
the $K_{\mathrm S}$ vs. $J-K_{\mathrm S}$ 
CMD in Figure \ref{fig:slopes2JK}, 
and 
the $H$ vs. $J-H$ 
CMD in Figure \ref{fig:slopes2JH}.  
Each diagram shows definitely smaller scatter; 
the slope of each fit and the standard deviation of the data points from the fit is 
shown
in Table \ref{tab:slope2}.  
For instance, 
the error in the slope in 
the $K_{\mathrm S}$ vs. $H-K_{\mathrm S}$
CMD 
based on all the 566 subfields is 0.03, but the deviation of each subfield point is 
as large as 0.077.  
In contrast, if the area is limited to the tile b349, 
the estimated error in the slope 
is increased to 0.21 due to the limited $H-K_{\mathrm S}$ range 
(compare Figure 3 (left) and Figure 4 (upper left corner)), 
but the deviation is reduced to 0.033.  
Thus, if we calculate the average slope for all the 566 subfields, 
we get seemingly smaller error estimates in the slopes.  
However, these subfields are likely to be inhomogeneous 
and should be analyzed separately.  

There seems to be real variation in the sight lines. 
In fact, it is often suggested that the extinction laws
show variations depending on sightlines  
(N06; 
\cite{gos09}; \cite{nat16}).  
The variation of the extinction law parameter $\alpha$, 
when fitted by a power law $A(\lambda) \propto \lambda^{-\alpha}$, 
is 
seen on scales as small as 5 arcsec \citep{gos09}.  
We will examine whether variation 
in the tile size of $\sim$ a degree really exists
or it 
is
due to some systematic observation effects, 
in the next section.  
Here, we note that the weighted-mean slopes of the 
$A(K_{\mathrm S}) / E(H-K_{\mathrm S}) $ and
$A(H) / E(J-H) $ 
CMDs are 
more consistent with those obtained by 
N06.  
In particular, 
the $A(K_{\mathrm S}) / E(H-K_{\mathrm S}) $ slope, 
which is derived from the maximum 
number of
stars, 
perfectly
agrees with 
N06.  


%
%
%

\begin{figure}
  \begin{center}
    \FigureFile(75mm,150mm){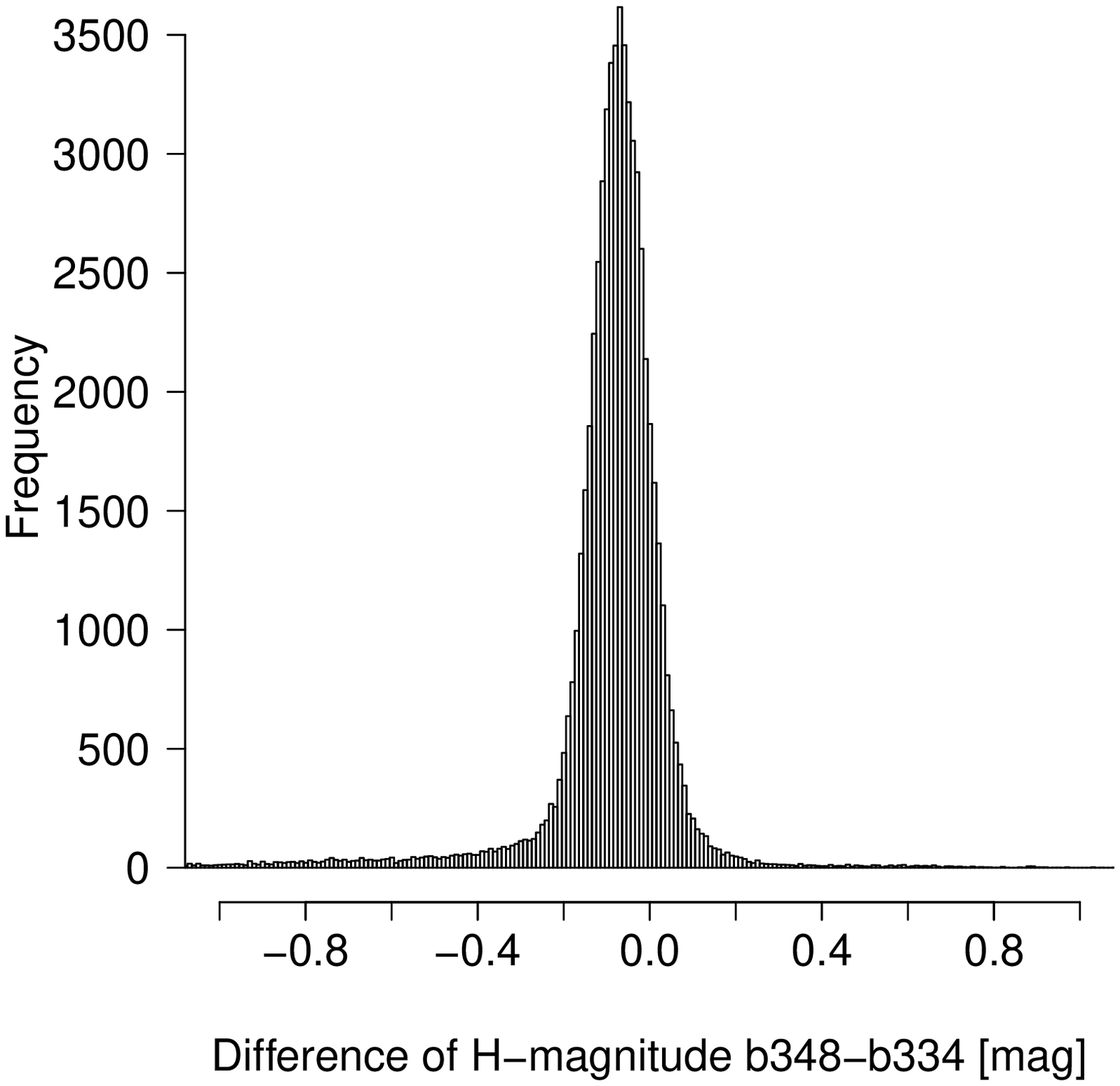}
    \FigureFile(75mm,150mm){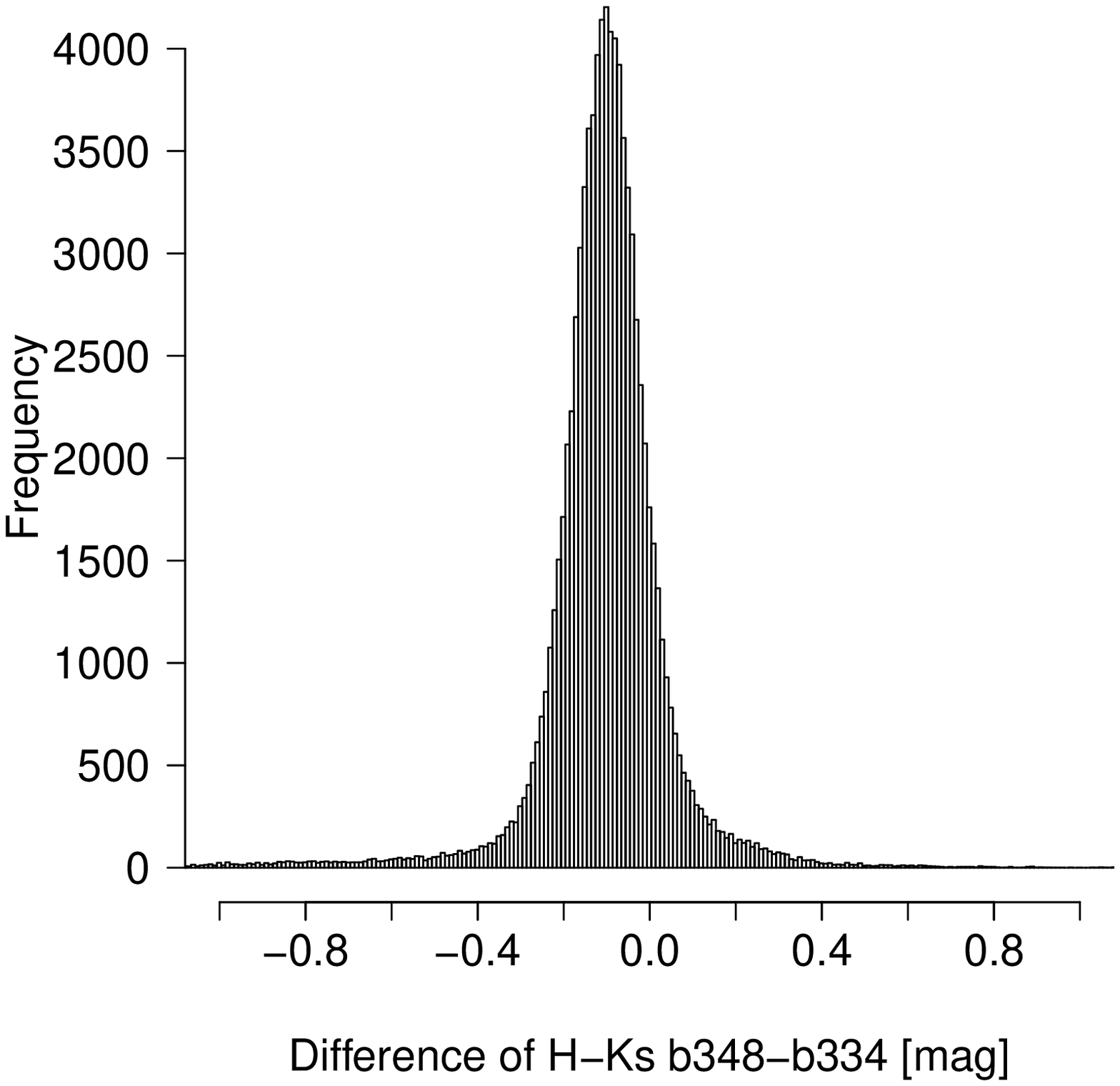}
  \end{center}
\caption{ Difference of $H$ (left)  and $H-K_{\mathrm S}$ (right) of common stars 
 between the adjacent b334 and b348 tiles. Each bin is 0.01 mag wide. 
}
\label{fig:tile-discr}
\end{figure}

\newpage
\begin{table}
\caption{Slopes of the 
$K_\mathrm{S}$ vs. $H-K_\mathrm{S}$,
$K_\mathrm{S}$ vs. $J-K_\mathrm{S}$, and
$H$ vs. $J-H$ 
CMDs and deviation from the fits.
}
\begin{tabular}{l|rll|rll|rll}
Tile & \#Subfields & $H-K_\mathrm{S}$\, Slope & {\scriptsize Deviation} & \#Subfields & $J-K_\mathrm{S}$\, Slope & {\scriptsize Deviation} & \#Subfields & $J-H$\, Slope & {\scriptsize Deviation} \\ \hline
 $b > 0.7\arcdeg$ & & & & & & & & & \\
 b349 &  54 & 1.58  $\pm$ 0.21 & 0.033 & 34 & 0.416 $\pm$ 0.05 & 0.033 & 34 & 1.22  $\pm$ 0.08 & 0.026\\
 b348 &  60 & 1.60  $\pm$ 0.11 & 0.029 & 33 & 0.504 $\pm$ 0.08 & 0.045 & 32 & 1.21  $\pm$ 0.14 & 0.035\\
 b347 &  67 & 1.27  $\pm$ 0.11 & 0.035 & 37 & 0.344 $\pm$ 0.07 & 0.035 & 37 & 1.20  $\pm$ 0.14 & 0.034\\
 b346 &  24 & 1.22  $\pm$ 0.28 & 0.044 &  9 & 0.446 $\pm$ 0.20 & 0.044 &  8 & 1.41  $\pm$ 0.39 & 0.017\\
 $ 0.7\arcdeg > b > 0\arcdeg$ & & & & & & & & & \\
 b335 &   3 & 2.12  $\pm$ 0.31 & 0.006 & 3  & 0.567 $\pm$ 0.29 & 0.031 &  3 & 0.964  $\pm$ 0.33 & 0.016\\
 b334 &  42 & 1.43  $\pm$ 0.19 & 0.048 & 0 & \ \ ---   &  &  2 & (1.07)  & \\
 b333 &  12 & 1.22  $\pm$ 0.39 & 0.037 & 0 & \ \ --- &      &   0 & \ \ ---   & \\
 b332 &  17 & 3.45  $\pm$ 3.00 & 0.046 & 2 & (0.969) & & 2 & (4.99) & \\  \hline
 $ 0\arcdeg > b$ & & & & & & & & & \\
 b321 &  27 & 1.24  $\pm$ 0.22 & 0.045 & 3  & 0.765 $\pm$ 0.10 & 0.007 &  2 & ($-$0.584) & \\
 b320 & 109 & 1.36  $\pm$ 0.08 & 0.046 & 75 & 0.425 $\pm$ 0.04 & 0.058 & 77 & 1.38  $\pm$ 0.07 & 0.035\\
 b319 &  66 & 1.48  $\pm$ 0.13 & 0.067 & 31 & 0.429 $\pm$ 0.05 & 0.064 & 32 & 1.19  $\pm$ 0.06 & 0.033\\
 b318 &  85 & 1.46  $\pm$ 0.05 & 0.037 & 0 & \ \ --- & & 0 & \ \ --- & \\ \hline
 (all & 566 & 1.30  $\pm$ 0.03 & 0.077 & 226 & 0.482 $\pm$ 0.02 & 0.074 & 227  & 1.72  $\pm$ 0.07 & 0.078)\\ \hline
 weighted mean & &  1.44  $\pm$ 0.04 & & & 0.423 $\pm$ 0.024 & & &  1.25 $\pm$ 0.04 &  \\ \hline \hline
 N06 Slope & &  1.44  $\pm$ 0.01  & & & 0.494 $\pm$ 0.006 & & &  1.42 $\pm$ 0.02 &  \\ 
\end{tabular}
\label{tab:slope2}
\end{table}

\begin{table}
\caption{Slopes of the 
$K_\mathrm{S}$ vs. $H-K_\mathrm{S}$,
$K_\mathrm{S}$ vs. $J-K_\mathrm{S}$, and
$H$ vs. $J-H$ 
CMDs in the  $ 0\arcdeg > b$ regions
when divided into $ 0\arcdeg > b > -0.7\arcdeg$ and $ -0.7 \arcdeg > b$.
}
\begin{tabular}{l|rl|rl|rl}
Tile & \#Subfields & $H-K_\mathrm{S}$\, Slope &
\#Subfields & $J-K_\mathrm{S}$\, Slope &
\#Subfields & $J-H$\, Slope 
\\ \hline
$ 0\arcdeg > b > -0.7\arcdeg$ &  &  & & & \\
 b321 &   5 & 1.13  $\pm$ 0.39  &0 &\ \ --- &0 &\ \ ---\\
 b320 &  11 & 1.54  $\pm$ 0.57 &3 &$-$0.33 $\pm$0.10&3&0.51$\pm$0.76 \\
 b319 &  1 & \ \ --- &0 &\ \ ---   &0&\ \ --- \\
 b318 &  2 & (9.05)&0&\ \ --- &0&\ \ ---  \\
$ -0.7 \arcdeg > b$ &  & & & & \\
 b321 &  22 & 1.40  $\pm$ 0.26 &3 &0.77  $\pm$ 0.10   &.2&($-$0.58) \\
 b320 &  98 & 1.31  $\pm$ 0.08 &72 &0.427 $\pm$0.045&74&1.42$\pm$0.07 \\
 b319 &  65 & 1.50  $\pm$ 0.14 &31 &0.429 $\pm$0.049&32&1.19$\pm$0.06 \\
 b318 &  83 & 1.49  $\pm$ 0.06 &0&\ \ --- &0&\ \ ---  \\ 
\end{tabular}
\label{tab:slope3}
\end{table}

\section{Discussion}
%
%
In the positive Galactic latitude, the slope is smaller at the 
tiles b333 and b334 closer to the Galactic plane, 
so we divided the four 
tiles at the negative Galactic latitude of
$|b|=0.7\arcdeg$ and 
listed the slopes of the resultant eight regions in Table \ref{tab:slope3}.  
The same trend as the positive Galactic latitude might exist in the negative latitude, but it does not seem significant.  
Rather, the slope seems to show variations in the VVV tile to VVV tile.  
This might not be surprising because, in the Vista photometry, 
after the zero point of the detector is calibrated by the Cambridge
Astronomical Survey Unit (CASU) procedure, 
the transformations between 
the VISTA and 2MASS systems 
are determined on a tile by tile basis, and 
there seems to be an appreciable amount of scatter in them \citep{sot13}; 
this is equivalent to changing the calibration for each tile 
(see also the CASU website\footnote{http://casu.ast.cam.ac.uk/surveys-projects/vista/
technical/photometric-properties}).   
We compared the magnitudes of the same stars detected 
in adjacent VVV tiles, and found small biases.  
An example is shown in Figure 7; 
the difference of $H$ magnitudes peaks at $-0.07$ and $H-K_{\mathrm S}$ peaks at $-0.11$  between the tiles b334 and b348.  
The influence of such differences is difficult to estimate.  
For instance, if all the magnitudes are shifted in one tile compared to another tile, 
that will not 
alter the slope in each tile.
However, these differences might explain at least part of the 
tile-to-tile slope variations.  

%
If we fit the 
$A(J), A(H)$, and $A(K_{\mathrm S})$ by a simple power law 
$A(\lambda) \propto \lambda^{-\alpha}$, 
the overall power law index is $\alpha \sim 2.2$, 
in contrast to the gentle power law $\sim \lambda^{-1.6}$ 
assumed in \citet{car89}.  
Such steep decline of the extinction as this work toward the longer wavelength 
is also observed for 
the Galactic center 
region (\cite{gos09}; \cite{nog18}), 
the two globular clusters at $l \sim 4\arcdeg$ and 
$l \sim 10\arcdeg$ \citep{alo15}, 
and for $\sim 4\,\mathrm{deg}^2$ area 
of the inner Milky Way \citep{nat16}.  
Therefore, a steep extinction law seems to be established very well 
in the central region of the Milky Way Galaxy.  
Furthermore, we also note that 
the $H$, $K_{\mathrm S}$, and $3.6 \mu$m extinction toward 
a distant Cepheid at $l \sim 30\arcdeg$ and 
probably another one at $l \sim 20\arcdeg$ 
\citep{tan17}, 
the extinction toward the cluster Westerlund 1 at $l \sim -20\arcdeg$ 
\citep{dam17}, 
and the infrared extinction ratios to a variety of objects 
in wider range of the galactic longitudes $l$ 
($5\arcdeg < l < 30\arcdeg$ in \cite{gon14}; 
 between $l \sim 27\arcdeg$ and $l \sim 100\arcdeg$ in \cite{ste09}; 
 whole inner Galactic disk in \cite{maj16})  
point to a steep extinction law also in somewhat outer regions.   


%
The variable extinction method determines 
the dependence of the variation in extinction on the variation in reddening such as 
$d A(K_{\mathrm S}) / d E(H-K_{\mathrm S})$.  
However, \citet{nat13} claimed that 
$A(I) / E(V-I) \neq d A(I) /d E(V-I)$ 
from their derivation of the extinction to the Galactic bulge 
using their $V$ and $I$ photometry from OGLE-III observations.  
They were 
surprised
by this result, but reasoned that 
since the fields in question that are apart a certain angle 
diverge linearly with distance, the variation is an average of different extinction laws, 
weighted 
strongly by 
the distant 
locations. 
They called this ``composite extinction bias''.
We examine if such bias exists in our data.  
We use the intrinsic luminosity parameters of the RC in \citet{nat13} and \citet{nat16}, and 
examine if the lines in the color magnitude diagrams (Figure 3) pass near the intrinsic 
points (the upper left blue squares).  
Extrapolating the reddening laws back to $E(H-K_{\mathrm S})=0$ approximately leads to 
a reasonable $K_{\mathrm S}$-band intrinsic magnitude of $K_{\mathrm S, 0}=12.9$; 
most of the acceptable least-squares fits (solid pink lines) pass 
near the intrinsic magnitude (within $\sim 0.1$ mag from blue squares).  
The other 
$J-K_{\mathrm S}$ vs. $K_{\mathrm S}$ and 
$J-H$ vs. $H$ CMDs 
show  similar results.  

In the fields examined by \citet{nat13} of $|b| \approx 5\arcdeg$, 
the 
component 
of extinction 
that contributes differentially to different sight lines 
and the kind of extinction that contributes equally to both sight lines 
(contributing to $A(I) / E(V-I)$ but not to $d A(I) /d E(V-I)$) 
are probably of similar strength. 
Thus it might lead to ``an unphysical difference in the value of 
$I_{\mathrm 0} = 0.24$ mag'' \citep{nat13}.  
In our fields of $|b| \approx 1\arcdeg$, however, 
the main contributor exists in the distant part, 
and its various parts are nearly equally apart.  
Therefore, $A(K_{\mathrm S}) / E(H-K_{\mathrm S}) \simeq d A(K_{\mathrm S}) / d E(H-K_{\mathrm S})$ seems to hold.  

%
Since a new PSF photometry of the VVV images is available \citep{alo18}
and provides more detection in the most crowded regions surveyed by VVV, 
we examined the PSF photometry magnitudes of the stars 
in the aperture photometry catalog.  
In all the area of this study, the aperture photometry catalog ({\tt vvvSource}) and the PSF photometry catalog ({\tt vvvPsfDophotZYJHKsSource}) were compared.  
 We used only the entries
whose magnitude errors are less than 0.1 mag 
in both the catalogs and one-to-one matches are guaranteed.  
The comparison was made tile by tile.  
It is not surprising that the agreement of magnitudes is generally very good 
because the PSF photometry also 
``relies heavily on the astrometric
and photometric solutions provided by CASU'' \citep{alo18}; 
it is calibrated with the aperture photometry and brought to the 
VISTA photometric system. 
However, the medians of differences in some corresponding tiles 
was found to exceed 0.05 mag.  
Therefore, we have estimated the influence of such magnitude differences to 
the resultant slopes in the color magnitude diagrams;  
the influence seems to be insignificant.  
Thus, our results are likely to hold for the new VVV PSF photometry also.  

Then, what caused the difference in the extinction law derived here and those in 
\citet{alo17}, 
who made use of 
PSF photometry 
of VVV data?  
Their measured ratios of total to selective extinction were
$A(K_{\mathrm S}) / E(H-K_{\mathrm S}) = 1.104 \pm 0.022 \pm 0.2$, 
which is significantly different from our ratio 1.44,  
and 
$A(K_{\mathrm S}) / E(J-K_{\mathrm S}) = 0.428 \pm 0.005 \pm 0.04$, 
which is rather similar to ours 0.423.  
The first cause of the difference would be that they use the VVV data obtained at smaller Galactic latitudes,  
thanks to the deeper PSF photometry.  
As shown above and in Figure 3 of \citet{alo17}, 
the extinction laws show variations depending on sightlines.  
If the ratio $A(K_{\mathrm S}) / E(H-K_{\mathrm S})$ is smaller 
near the Galactic plane, it might explain the difference.  
Another possibility is the slight differences in analysis.  
\citet{alo17} used the whole area CMDs.  
They divided them into narrow sections of color, 
generated histograms of the stars in each color bins, and 
fitted the histograms with a second-order polynomial function 
plus two Gaussians representing the RC and other stars. 
The RC stars seem to be separated well, and 
the Gaussian fitting results were used to derive their extinction law.  
In contrast, we have selected a rectangle region that is supposed to be occupied only by the RC stars in each CMD, 
taken from each subfield, and fit the magnitude and color distributions with a Gaussian function.  
We made such procedures to select only the RC stars in the bulge, because 
other various components 
in the foreground and background (e.g., \cite{alo18} and \cite{gon18}) 
can contaminate.  
Furthermore, the tile-to-tile photometry difference found here might have 
complicated the whole area CMDs; 
our average slopes for all area CMDs are different from the mean values of each tile.  


\section{Summary}
%
%
The average of the ratios of total to selective extinction 
derived from the VVV aperture photometry of RC stars 
$A(K_{\mathrm S}) / E(H-K_{\mathrm S}) = 1.44 \pm 0.04 $,
$A(K_{\mathrm S}) / E(J-K_{\mathrm S}) = 0.423 \pm 0.024$,
$A(H)             / E(J-H)             = 1.25 \pm 0.04$ 
are similar to those obtained 
from RC stars with the IRSF/SIRIUS system in previous studies.  
This steep extinction law 
has been found to be 
also typical of
recent deep imaging towards the inner Galaxy. 
This extinction law
is steeper than the 
classical
ones \citep{rie85,car89} 
based on less deep observations, 
and is in line with recent results based on deep imaging surveys
\citep{nat16,dam17}.  
In
the near infrared, this implies smaller $A(K_\mathrm{S})$ and larger distances, 
which has a strong impact on
inner Galaxy studies.


\bigskip
This work was 
partly supported by the Grants-in-Aid
Scientific Research (C) 21540240, (A) 18H03719, 18H05441, the Global
COE Program ``The Next Generation of Physics, Spun from
Universality and Emergence'' from the Ministry of Education,
Culture, Sports, Science and Technology (MEXT) of Japan, and 
the ``UCHUGAKU'' project of the Unit of Synergetic Studies for Space, 
Kyoto University.  
Nishiyama acknowledges support by JSPS KAKENHI,
Grant-in-Aid for Challenging Exploratory Research 15K13463, 18K18760,
and Grant-in-Aid for Scientific Research (A) 19H00695.



\end{document}